\documentclass[preprint]{revtex4-1}
\pdfoutput=1
\usepackage[utf8]{inputenc}
\usepackage[english]{babel}
\usepackage{amssymb,amsfonts,amsmath}
\usepackage[stretch=10]{microtype}
\usepackage[titletoc,title]{appendix}
\usepackage{graphicx}
\graphicspath{{figures/}}

\usepackage{hyperref}
\hypersetup{
    bookmarks=true,
    unicode=false,
    pdftoolbar=true,
    pdfmenubar=true,
    pdffitwindow=false,
    pdfstartview={FitH},
    pdftitle={Bloch wave scattering on impurity in 1D Dirac comb model},
    pdfauthor={Stepan Botman},
    pdfsubject={Theoretical physics},
    pdfkeywords={Dirac comb} {resistivity} {Boltzmann equation} {scattering},
    colorlinks=true,
    linkcolor=blue,
    citecolor=green,
}

\begin{document}

\title{Bloch wave scattering on pseudopotential impurity in 1D Dirac comb model.}

\author{Stepan Botman}
\email{sbotman@innopark.kantiana.ru}
\affiliation{Faculty of physics and technology, Immanuel Kant Baltic Federal University}

\author{Sergey Leble}
\email{leble@mif.pg.gda.pl}
\affiliation{Faculty of physics and technology, Immanuel Kant Baltic Federal University}
\date{\today}

\begin{abstract}
This paper presents calculation of electron-impurity scattering coefficient of Bloch waves for one dimensional Dirac comb potential. The impurity is also modeled as delta function pseudopotential that allows explicit solution of Schr\"odinger equation and scattering problem for Bloch waves. 
% Obtained scattering probability exhibit reflectionless properties for some parameters.
\end{abstract}


\maketitle

\section{Introduction}

Scattering theory provide powerful direct ways for quantum systems treatment and and allow to obtain essential information about these systems. In Solid State Physics scattering theory has been used to describe various transport phenomena. The main feature in this case is presence of periodic potential due to the lattice structure which results in basis of Bloch functions.

The scattering in terms of Bloch functions has been studied in a number of works. The works of Morgan~\cite{Morgan66} and Newton~\cite{Newton91} 
are based on the Korringa, Kohn and Rostoker equation~\cite{korringa47,kohn54} and give rather a general description of Bloch electron scattering on impurities in crystals.

Unfortunately, the resulting expressions for the scattering amplitudes are complicated, its direct use in evaluation of transport properties is difficult. In this paper we propose, perhaps, the simplest model formulation and realization of the scattering problem on a base of Dirac comb potential and ZRP model for an impurity. It is known that the mean velocity of conductivity electrons is low in a sense that the electron de Broglio wavelength is large compared with atom dimension~\cite{DemkovZRP}. Its justify applications of zero-range potentials (ZRP). 

Dirac Comb model is a special case of the Kronig-Penney model~\cite{kronigpenney}, which ranks among small number of exact solvable problems in quantum mechanics.
It is interesting to investigate aspects of electron Bloch scattering within simple model, which allows to insight into basic properties of the Bloch electron scattering in systems with periodic potentials.

In this work we construct flux normalized Bloch wavefunctions and use them for impurity scattering probability determination.
In section~\ref{section:blochwf} we start from reproducing some classical results of electron properties in Dirac comb potential.
Then, in section~\ref{section:ef} we use obtained expression for energy dispersion to study chemical potential temperature dependence for considered system.
Next, in section~\ref{section:flux} flux normalisation for Bloch wave basis set is provided.
Finally, in section~\ref{section:impurity} the ZRP  impurity scattering probability is derived in explicit form.
Section~\ref{section:summary} contains summary and discussion.

\section{Dirac comb defect scattering}

\subsection{Bloch wavefunctions} \label{section:blochwf}

Let us reproduce classical calculation (see~\cite{Kittel-introSSP}) of electron properties in a lattice under the influence of a weak attractive Dirac Comb potential for a reader convenience. In Dirac Comb model a single atom potential in each cell is reduced to a Dirac delta function.

% Solvable models in Quantum Mechanics

Consider an electron moving in potential of equidistant Dirac delta functions:
\begin{equation}
\hat{V} = \beta \delta (x-na),
\; n = 0, \pm 1, \ldots
\label{diraccomb-potential}
\end{equation}
where \(\beta\) --- parameter of potential, \(a\) --- period of cell. Generally, the parameter \(\beta\) can be both negative and positive.

Each delta function in~(\ref{diraccomb-potential}) represents the simplest zero range potential. For such a case, Shr\"odinger equation for potential~(\ref{diraccomb-potential}) can be replaced with boundary condition imposed on wave function (for more detailed information see~\cite{DemkovZRP}):
\begin{equation}
\left. \psi' \vphantom{\int} \right|^{x=na+0}_{x=na-0} -
\left.\frac{2m}{\hbar^2} \beta \psi \right|_{x=na} =0,
\; n = 0, \pm 1, \ldots
\label{1d-zrp-boundary}
\end{equation}
where \(\psi\) is wavefunction. 
Obviously, first term of (\ref{1d-zrp-boundary}) is non zero and wavefunction \(\psi\) have finite discontinuity of first derivative at points of potential location.

For one dimensional problem Bloch wavefunction must be eigenfunction of both Hamilton and translation operator.
Wave function for the domain between two point potentials located at \(x=0\) and \(x=a\) (i.e. for \(x \in \left[0, a\right) \) ) should be linear combination of two plain waves:
\begin{equation}
\psi_{\left[0, a\right)} = 
A e^{i k x} + B e^{-i k x},
\label{wf-free-0}
\end{equation}
where \(A\), \(B\) --- two complex constants, \(k=\sqrt{2 m E}/\hbar\) --- momentum of electron.

According to the definition the shift operator is given by:
\begin{equation}
\hat{T}_a \psi (x) = \psi(x+a).\
\label{shift-operator}
\end{equation}

Corresponding eigenvalue problem for  ~(\ref{shift-operator}) takes the form:
\begin{equation}
\hat{T}_a \psi (x) = \mu \psi.
\label{shift-operator-eigen}
\end{equation}

We solve the problem ~(\ref{shift-operator-eigen}) demanding finiteness of wavefunction \(\psi\) at both infinities:
\begin{equation}
\mu \psi(x) = \psi(x+a)
\; \rightarrow \;
\left| \mu \right| = 1
\; \rightarrow \;
\mu = e^{i K a}.
\label{shift-operator-eigen-K}
\end{equation}
where \(K\) --- real constant. Addition limitation on \(K\) can be applied in case of Born-Karmann condition (\(N\) cell ring-like sewing) is assumed. In this case:
\begin{equation}
K = \frac{2 \pi n}{N a},
\; n = 0, \pm 1, \ldots
\end{equation}

Applying (\ref{shift-operator-eigen}) to (\ref{wf-free-0}) and taking into account~(\ref{shift-operator-eigen-K}) one may get wavefunction for the domain~\(\left[-a,0\right)\):
\begin{equation}
\psi_{\left[-a, 0\right)} = 
e^{i K a} \left(
A e^{i k (x+a)} + B e^{-i k (x+a)}
\right).
\label{wf-free-0-shifted}
\end{equation}

\begin{figure*}[t]
\includegraphics[width=\textwidth]{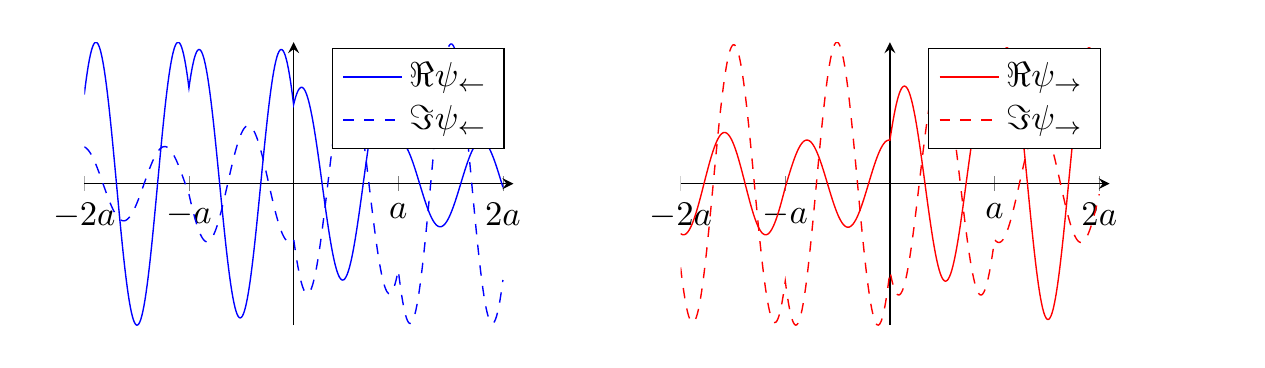}
\caption
{Left and right Bloch functions (real and imaginary parts are labeled with continuous and dashed line respectively).}
\end{figure*}

%\begin{figure*}[tbh]
%\begin{minipage}{0.47\textwidth}
%\begin{tikzpicture}
%\begin{axis}[axis lines=middle, xtick={-2,-1,1,2}, xticklabels={\(-2a\),\(-a\),\(a\),\(2a\)}, ytick=\empty, xmin=-2, xmax=2.1, width=\columnwidth, height=0.75\columnwidth]
%\addplot [color=blue, style=solid] table [x index={0},y index={1}]{ppblochwf.dat};
%\addlegendentry{\( | \phi_{\leftarrow} | \)};
%\addplot [color=blue, style=dashed] table [x index={0},y index={2}]{ppblochwf.dat};
%\addlegendentry{\( \text{arg} ( \phi_{\leftarrow} ) \)};
%\end{axis}
%\end{tikzpicture}
%\end{minipage}
%\hfill
%\begin{minipage}{0.47\textwidth}
%\begin{tikzpicture}
%\begin{axis}[axis lines=middle, xtick={-2,-1,1,2}, xticklabels={\(-2a\),\(-a\),\(a\),\(2a\)}, ytick=\empty, xmin=-2, xmax=2.1, , width=\columnwidth, height=0.75\columnwidth]
%\addplot [color=red, style=solid] table [x index={0},y index={3}]{ppblochwf.dat};
%\addlegendentry{\( | \phi_{\rightarrow} | \)};
%\addplot [color=red, style=dashed] table [x index={0},y index={4}]{ppblochwf.dat};
%\addlegendentry{\( \text{arg} ( \phi_{\rightarrow} ) \)};
%\end{axis}
%\end{tikzpicture}
%\end{minipage}
%\caption{Periodic part of left and right Bloch functions \(\psi_{\rightarrow} = e^{iKx} \phi_{\rightarrow}\) (real and imaginary parts).}
%\end{figure*}

Substitution of~(\ref{wf-free-0}) and~(\ref{wf-free-0-shifted}) into continuity condition for \(\psi\) at point \(x=0\) and~(\ref{1d-zrp-boundary}) gives us a system:
\begin{widetext}
\begin{subequations}
\begin{equation}
A \left( e^{iKa}e^{ika} - 1 \right) +
B \left( e^{iKa}e^{-ika} - 1 \right) = 0,
\label{diraccomb-AB-1}
\end{equation}
\begin{equation}
A \left( ik - ik e^{iKa}e^{ika} - \frac{2m}{\hbar^2} \beta \right) +
B \left( -ik + ik e^{iKa}e^{-ika} - \frac{2m}{\hbar^2} \beta \right) = 0.
\label{diraccomb-AB-2}
\end{equation}
\end{subequations}
\end{widetext}

System of equations (\ref{diraccomb-AB-1}) and~(\ref{diraccomb-AB-2}) is system of linear homogeneous equations for \(A\) and \(B\). It have one trivial solution and may have one fundamental solution. Condition of equations consistency (determinant of coefficients matrix equals zero) gives us:
\begin{equation}
\cos (Ka) = \cos (ka) +
\frac{m \beta}{\hbar^2} \frac{\sin (ka)}{k}.
\label{diraccomb-dispersion}
\end{equation}

\begin{figure*}[hp]
\centering
\includegraphics[width=\textwidth]{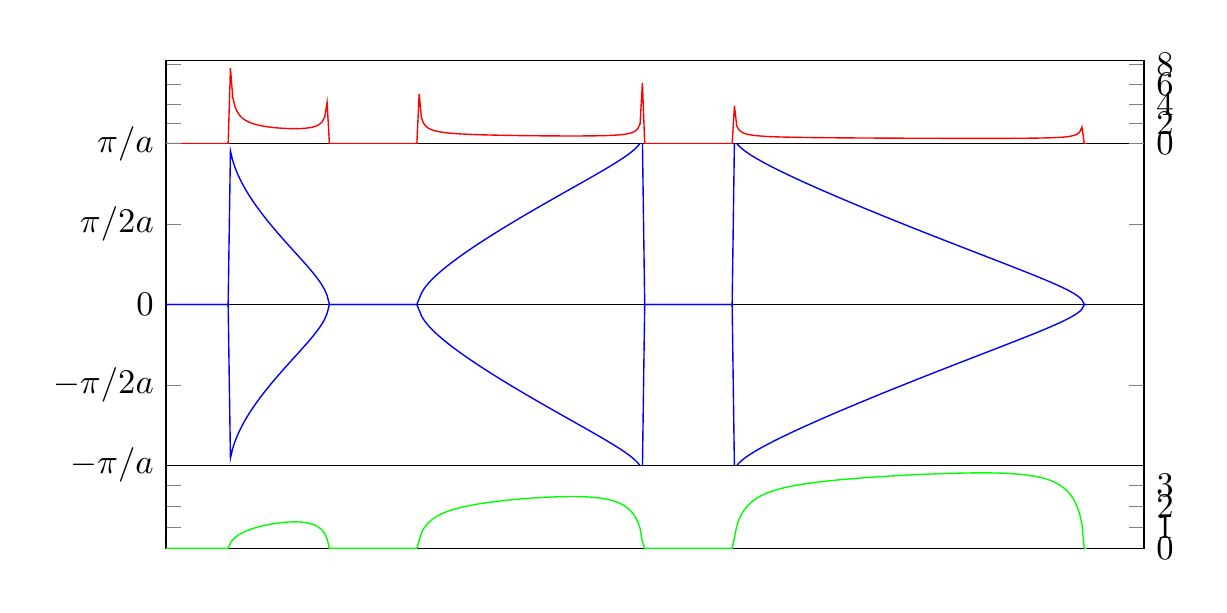}
\caption{Density of states (top, red), energy dispersion (middle, blue), electron velocity (bottom, green).}
\label{edos}
\end{figure*}

\begin{figure*}[hp]
\centering
\includegraphics[width=\textwidth]{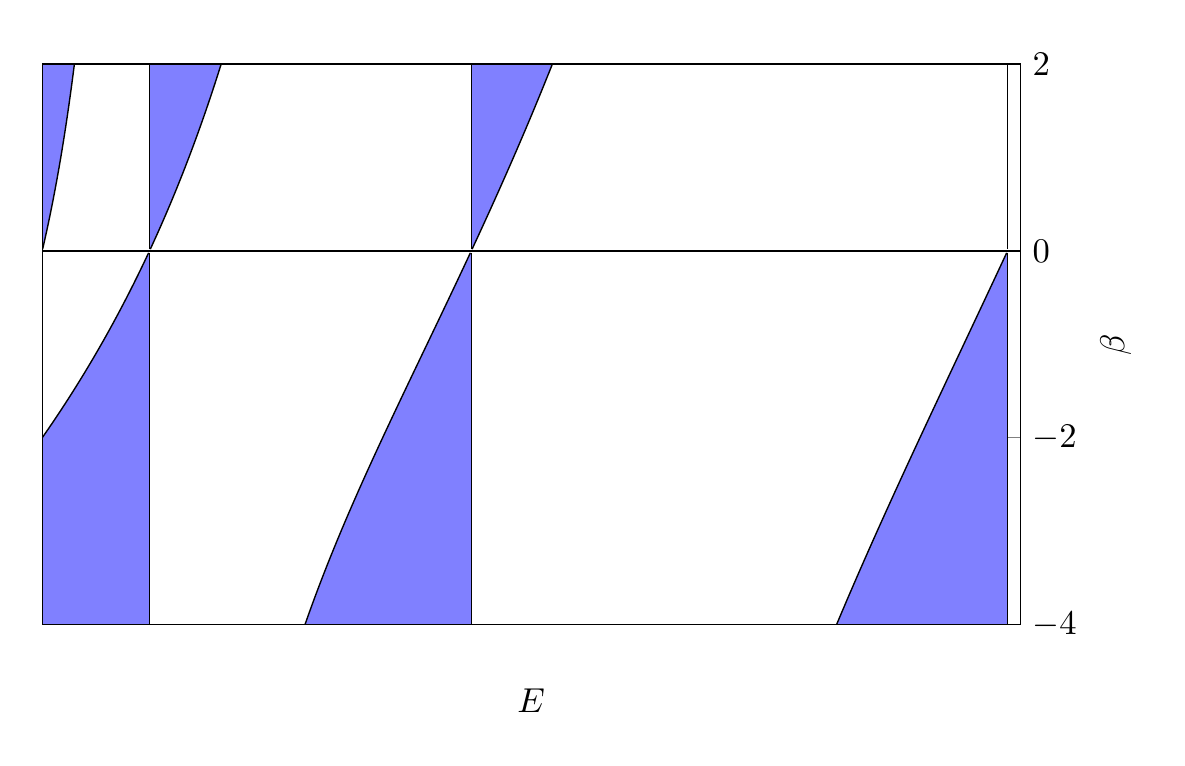}<
\caption{Zone edges depending on \(\beta\) parameter (parameter \(a=1\)).}
\label{zoneshifts}
\end{figure*}

Equation (\ref{diraccomb-dispersion}) is well known and in fact gives a band structure --- dependence between energy \(E = \hbar k^2 / 2m\) and quasi wave vector \(K\). Typical band structure of Dirac comb is shown in figure~\ref{edos}. Tuning \(\beta\) and \(a\) parameters allows to set arbitrary zone/bandgap ratio (as shown in figure~\ref{zoneshifts}) which can be useful for building real systems approximations.

Moreover, (\ref{diraccomb-dispersion}) allow to derive analytical expressions for several solid state physics quantities such as density of states and electron velocity (see~\cite{Kittel-introSSP}):
\begin{gather}
\rho (E) = \pm \frac{2}{\pi} \left( \frac{d E}{d K} \right)^{-1} = 
\frac{2}{\pi} K'(E), 
\label{dos-analytical}
\\
v (E) = \frac{1}{\hbar} \frac{d E}{d K} =
\frac{1}{\hbar} \frac{1}{K'(E)}.
\label{velocity-analytical}
\end{gather}

Expressing \(K(E)\) from (\ref{diraccomb-dispersion}):
\begin{equation}
K(E) = \frac{1}{a} \arccos \left[ 
\cos \left(\frac{\sqrt{2mE}}{\hbar} a \right) \right.
+ \left. \frac{m \beta}{\hbar \sqrt{2mE}} 
\sin \left(\frac{\sqrt{2mE}}{\hbar} a \right)
\right],
\label{KE}
\end{equation}
evaluating derivate of~(\ref{KE}) one may obtain analytical form of~(\ref{dos-analytical}) and~(\ref{velocity-analytical}).

It's essential to note that for every allowed energy we will have two Bloch functions: for \(K\) and \(-K\), which are associated with left and right Bloch waves.

\subsection{Fermi energy} \label{section:ef}

Obtained expression~(\ref{dos-analytical}) for density of states allows us to calculate Fermi energy and chemical potential for electron in Dirac comb potential.
We start from Fermi-Dirac distribution:
\begin{equation}
n_{f} (E,T)= \frac{1}{e^{\frac{E-\mu(T)}{k_B T}} + 1},
\label{Fermi-distribution}
\end{equation}
where \(k_B\) --- Boltzman constant, \(\mu\) --- chemical potential ( \(\mu(T=0)=E_f\), \(E_f\) --- Fermi energy), \(T\)~---~temperature.

Obviously, (\ref{Fermi-distribution}) must satisfy normalization condition for a total number of electrons. We will take into account that at zero temperature Fermi distribution function reduces to a step function:
\begin{multline}
N_e =
\int \limits_0^{\infty} n_{f} (E,T=0) \rho (E) dE =
\int \limits_0^{E_f} \rho (E) dE = \\
= \left. \sum \limits_{\text{valence bands}} \frac{2 a}{\pi} K(E)
\right|_{E_{\text{band min}}}^{E_{\text{band max}}} +
\left. \frac{2 a}{\pi} K(E) \right|_{E_{\text{band min}}}^{E_f} =\\
= \sum \limits_{\text{valence bands}} 2 N +
\left. \frac{2 a}{\pi} K(E) \right|_{E_{\text{band min}}}^{E_f},
\label{Efermi-condition}
\end{multline}
where \(\rho(E)\) --- density of states, \(N_e\) --- total number of electrons, \(N\) --- number of cells in wire (assuming each atom in wire has only one electron).
Thus one may calculate \(E_f\) within this model using~(\ref{Efermi-condition}) and~(\ref{KE}).

Chemical potential can be obtained in similar way:
\begin{equation}
N_e = \int \limits_0^{\infty} n_{f} (E,T,\mu) \rho (E) dE =
\int \limits_0^{\infty} n_{f} (E,T,\mu) \frac{2 a}{\pi} K'(E) dE,
\label{chempot-condition}
\end{equation}

In some cases it may be more convenient to use the following form (here only valence band is taken into account):
\begin{multline}
N_e = \int \limits_0^{\infty} n_{f} (E,T,\mu) \frac{2}{\pi} K'(E) dE = \\
= \frac{2a}{\pi} \underbrace{n_{f}' (E,T,\mu)}_{\mbox{0 or 1}}
\underbrace{K (E)}_{\mbox{0 or \(\pi/a\)}}
\left. \vphantom{\int} \right|_{E_{\text{band min}}}^{E_{\text{band max}}}
- \frac{2 a}{\pi} \int \limits_{E_{\text{band min}}}^{E_{\text{band max}}}
n_{f}' (E,T,\mu) K(E) dE =\\
= 2 - \int \limits_{E_{\text{band min}}}^{E_{\text{band max}}}
n_{f}' (E,T,\mu) K(E) dE.
\label{chempot-norm}
\end{multline}

Chemical potential temperature dependence calculated within numerical procedure showed negligible shifts for points within band except the \(k_B T\)~neighborhood of upper bandedge, where local maximum is formed.

\subsection{Flux normalization} \label{section:flux}

It's essential to note that \(K\) in~(\ref{diraccomb-dispersion}) is in cosine so for every allowed energy we will have two Bloch functions: for \(K\) and \(-K\). Bloch functions for \(K\) and \(-K\) will give us appropriate flux in opposite directions.

For one dimension flux can be calculated as follows:
\begin{equation}
j (\psi) = -i  \frac{\hbar}{2m} \left(
\psi* \frac{d}{dx} \psi - \psi \frac{d}{dx} \psi*
\right).
\label{1d-flux}
\end{equation}

Direct substitution (\ref{wf-free-0}) into~(\ref{1d-flux}) yields:
\begin{equation}
j (\psi) = \frac{\hbar k}{m} \left( |A|^2 - |B|^2 \right).
\label{diraccomb-flux}
\end{equation}

Equation (\ref{diraccomb-flux}) is true for any \(x\).

Solution of system (\ref{diraccomb-AB-1},~\ref{diraccomb-AB-2}) gives dependence between coefficients \(A_\pm = f_\pm(k,K,\beta) B_\pm\) and furthermore several equivalent form of solution can be obtained:
\begin{equation}
A_\pm = B_\pm \left[ 1 + i \frac{4m}{\hbar^2} \frac{k}{\beta}
\left( e^{\pm iKa} e^{-ika} \right) \right],
\label{ab-1}
\end{equation}
or
\begin{equation}
B_\pm = A_\pm \frac{e^{\pm iKa}e^{ika} -1}{e^{\pm iKa}e^{-ika} -1}.
\label{ab-2}
\end{equation}
From now on we will use~(\ref{ab-2}) as its form is more convenient.

Let us introduce the following coefficients \(b_+\) and \(b_-\) for shortening:
\begin{equation}
A_{\pm} = B_{\pm} b_{\pm}, \quad
b_{\pm} = \frac{e^{\pm iKa}e^{-ika} -1}{e^{\pm iKa}e^{ika} -1}.
\label{bpm}
\end{equation}

Substitution of (\ref{ab-2}) into~(\ref{diraccomb-flux}) gives:
\begin{equation}
j(\psi_{\rightleftarrows}) = \frac{\hbar k}{m} |A_{\pm}|^2
\frac{\pm \sin(Ka) \sin(ka)}{1-\cos(\pm Ka +ka)}.
\label{ab-2-flux}
\end{equation}

Next, equation~(\ref{ab-2-flux}) is used for flux normalization:
\begin{equation}
j(\psi_{\rightleftarrows}) = \pm 1.
\label{fux-normalization}
\end{equation}

Left and right waves are normalized separately. Equations~(\ref{fux-normalization}),~(\ref{ab-2-flux}) and~(\ref{ab-2}) are used to determine coefficient \(A_{\pm}\) and \(B_{\pm}\) consequently.

\subsection{Impurity scattering} \label{section:impurity}

Generally Bloch wavefunction for Dirac comb potential~(\ref{diraccomb-potential}) has the following form:
\begin{equation}
\psi_{\rightleftarrows} = e^{\pm iKna } \left(
A_{\pm} e^{i k x} + B_{\pm} e^{-i k x} \right) \;
\text{for} \; x \in \left[na, (n+1)a \right).
\end{equation}

Let us assume that electron Bloch wave propagate from~\(-\infty\) to~\(\infty\) and scatter at potential center \( \gamma \delta(x-x_0) \) located at point \(x_0 \in \left(ma, (m+1)a \right)\).
Thus, scattering ansatz appears as follows: 
\begin{align}
\Psi_{\left(ma, x_0\right]} &= C_i \psi_{B+} + C_s \psi_{B-},
\\
\Psi_{\left[x_0, (m+1)a\right)} &= C_t \psi_{B+},
\end{align}
where index \(i\) stands for incident, \(s\) stands for scattered, \(t\) stands for transmitted wave.

Now boundary condition (analogue of~(\ref{1d-zrp-boundary})) should be applied at point \(x_0\), that gives:
\begin{multline}
C_t e^{iKma} ik \left( A_+ e^{ikx_0} - B_+ e^{-ikx_0} \right) - \\
- C_i e^{iKma} ik \left( A_+ e^{ikx_0} - B_+ e^{-ikx_0} \right) - \\
- C_s e^{-iKma} ik \left( A_- e^{ikx_0} - B_- e^{-ikx_0} \right) = \\
= \frac{2m \gamma}{\hbar^2} C_t e^{iKma} \left( A_+ e^{ikx_0} + B_+ e^{-ikx_0} \right).
\label{1dx0-1}
\end{multline}

We should add one more condition for continuity at point \(x_0)\).
\begin{multline}
C_i e^{iKma} \left( A_+ e^{ikx_0} + B_+ e^{-ikx_0} \right) + \\
+ C_s e^{-iKma} \left( A_- e^{ikx_0} + B_- e^{-ikx_0} \right) = \\
C_t e^{iKma} \left( A_+ e^{ikx_0} + B_+ e^{-ikx_0} \right).
\label{1dx0-2}
\end{multline}

Both (\ref{1dx0-1}) and (\ref{1dx0-2}) form a system.
We state \(C_i=1\) (it correspond to incident flux normalized to one) and solve system for \(C_s\) and \(C_t\):

\begin{widetext}
\begin{gather}
C_s = - \frac{ \left( A_+ e^{ik x_0} + B_+ e^{-ik x_0} \right)^2 \gamma m}
{i \hbar^2 k \left( A_- B_+ - A_+ B_- \right) + \gamma  m \left( A_- B_+  + A_+ B_- \right) + A_- A_+ \gamma e^{2 ik x_0} m + B_- B_+ \gamma e^{-2 ik x_0} m}, 
\label{scattering-probability}\\
C_t = \frac{ i \hbar^2 k \left( A_- B_+ - A_+ B_- \right)}
{i \hbar^2 k \left( A_- B_+ - A_+ B_- \right) + \gamma  m \left( A_- B_+  + A_+ B_- \right) + A_- A_+ \gamma e^{2 ik x_0} m + B_- B_+ \gamma e^{-2 ik x_0} m}.
\end{gather}
\end{widetext}

One in shorter form, using~(\ref{bpm}):
\begin{widetext}
\begin{gather}
C_s = - \frac{1}{B_-}
\frac{ \left( b_+ e^{ik x_0} + e^{-ik x_0} \right)^2 \gamma m}
{( b_- - b_+ )( i \hbar^2 k + \gamma  m ) + \gamma m \left( b_- b_+ e^{2 ik x_0} + e^{-2 ik x_0} \right)}, 
\\
C_t = \frac{ i \hbar^2 k \left( b_- - b_+ \right)}
{( b_- - b_+ )( i \hbar^2 k + \gamma  m ) + \gamma m \left( b_- b_+ e^{2 ik x_0} + e^{-2 ik x_0} \right)}.
\end{gather}
\end{widetext}

\(W=\left| C_s \right|^2\) gives us scattering probability. \(T=\left| C_t \right|^2\) gives us transmission probability. 
In order to calculate scattering probability one should replace \(A_{\pm}\) and \(B_{\pm}\) with~(\ref{ab-2}) and~(\ref{ab-2-flux}) taking into account flux normalization~(\ref{fux-normalization}).

\begin{figure*}[p]
\centering
\includegraphics[width=\textwidth]{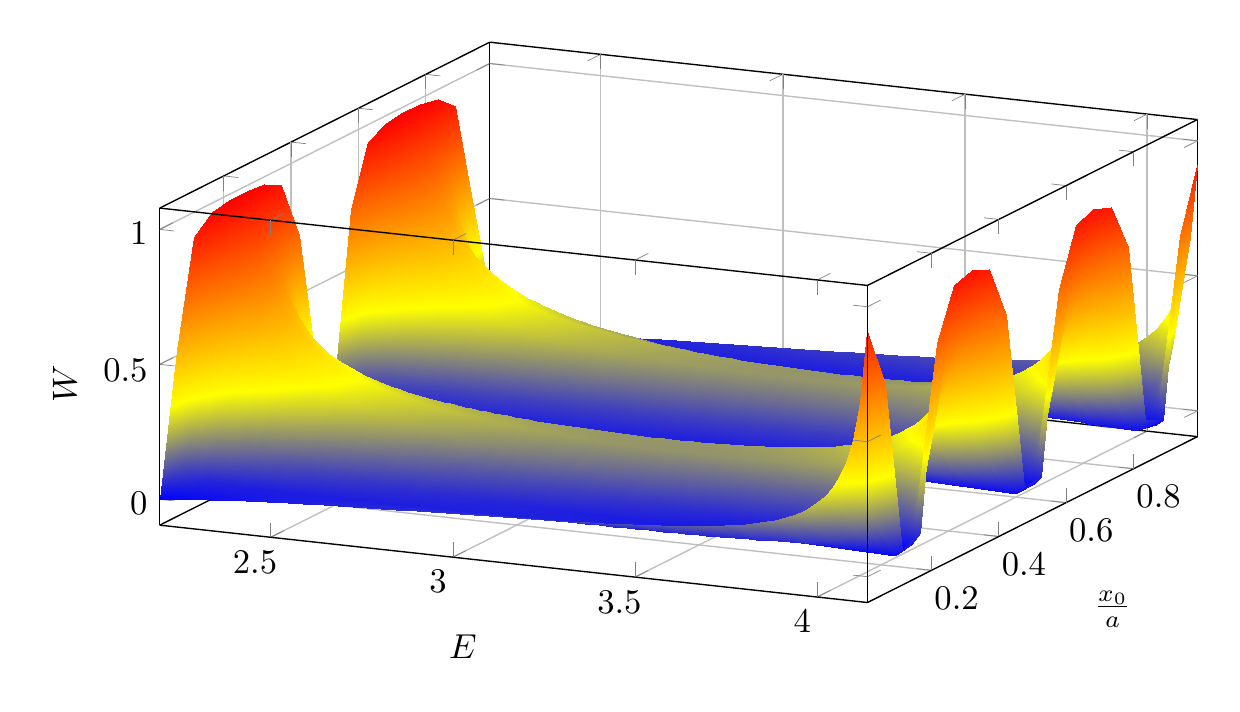}
\caption{Scattering probability \(W\) depending on scatterer position~\(x_0\).}
\end{figure*}

\begin{figure*}[hp]
\includegraphics[width=\textwidth]{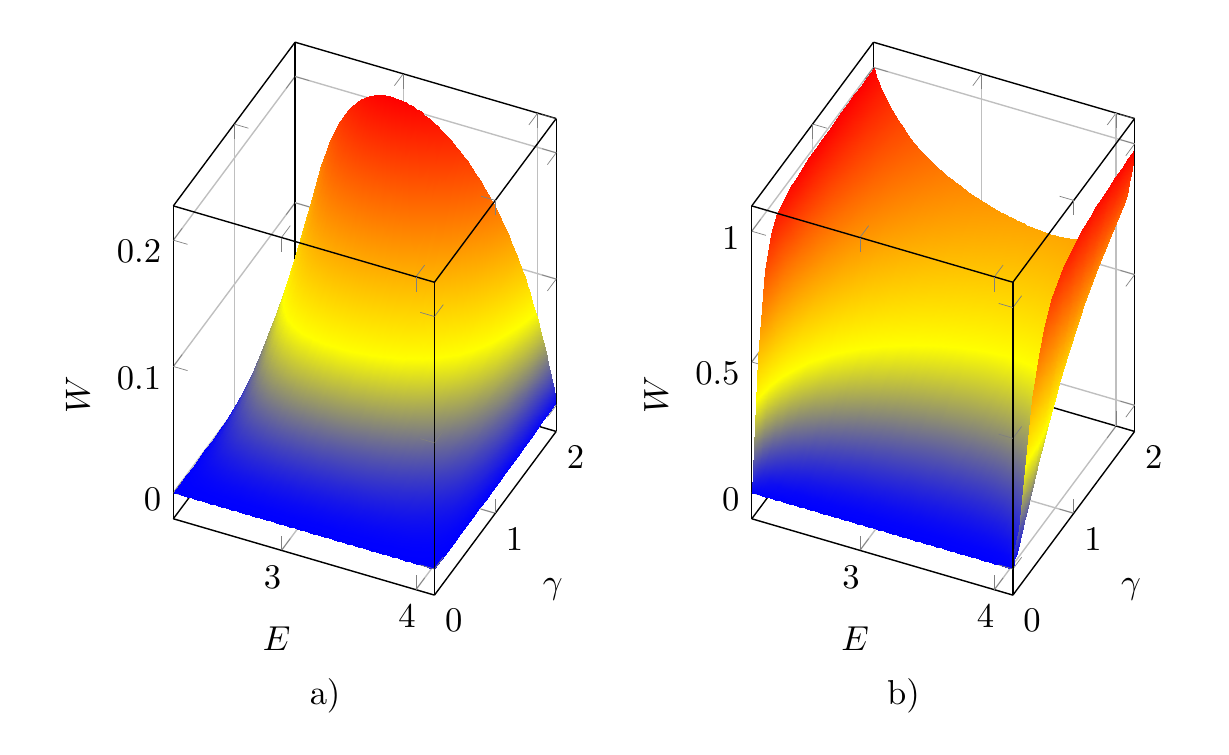}
\caption{Scattering probability \(W\) depending on scatterer strength~\(\gamma\), where a) --- scatterer located at \(x_0=\frac{a}{2}\) and b) --- scatterer located at \(x_0=\frac{a}{4}\) ).}

\end{figure*}

% \section{Case of dipole potential}
% Potential of the following form:
% \begin{equation}
% \hat{V}= \alpha \frac{d}{dx} \delta(x-x_0).
% \end{equation}

\section{Summary and discussing} \label{section:summary}
Scattering probability of Bloch electron on impurity depending on scatterer parameters and energy was obtained. Data showed nontrivial behaviour of scattering probability. It is planned to apply the results in the transport phenomena theory.

One dimensional scattering problem gives us a good "toy" model, that have very interesting application in the inverse scattering method of the soliton theory (nonlinear equation of KdV type)~\cite{novikov1984theory}. The results of scattering on ZRP theory may be applied in the context of integrable potentials theory as well~\cite{leble2014integrable}, its Bloch functions version could give an impulse to develop the theory.

\bibliography{references}
\bibliographystyle{unsrt}

\pagebreak

\begin{appendices}
\section{Resistivity calculation workflow}
\begin{figure}[th]
\includegraphics[width=\textwidth]{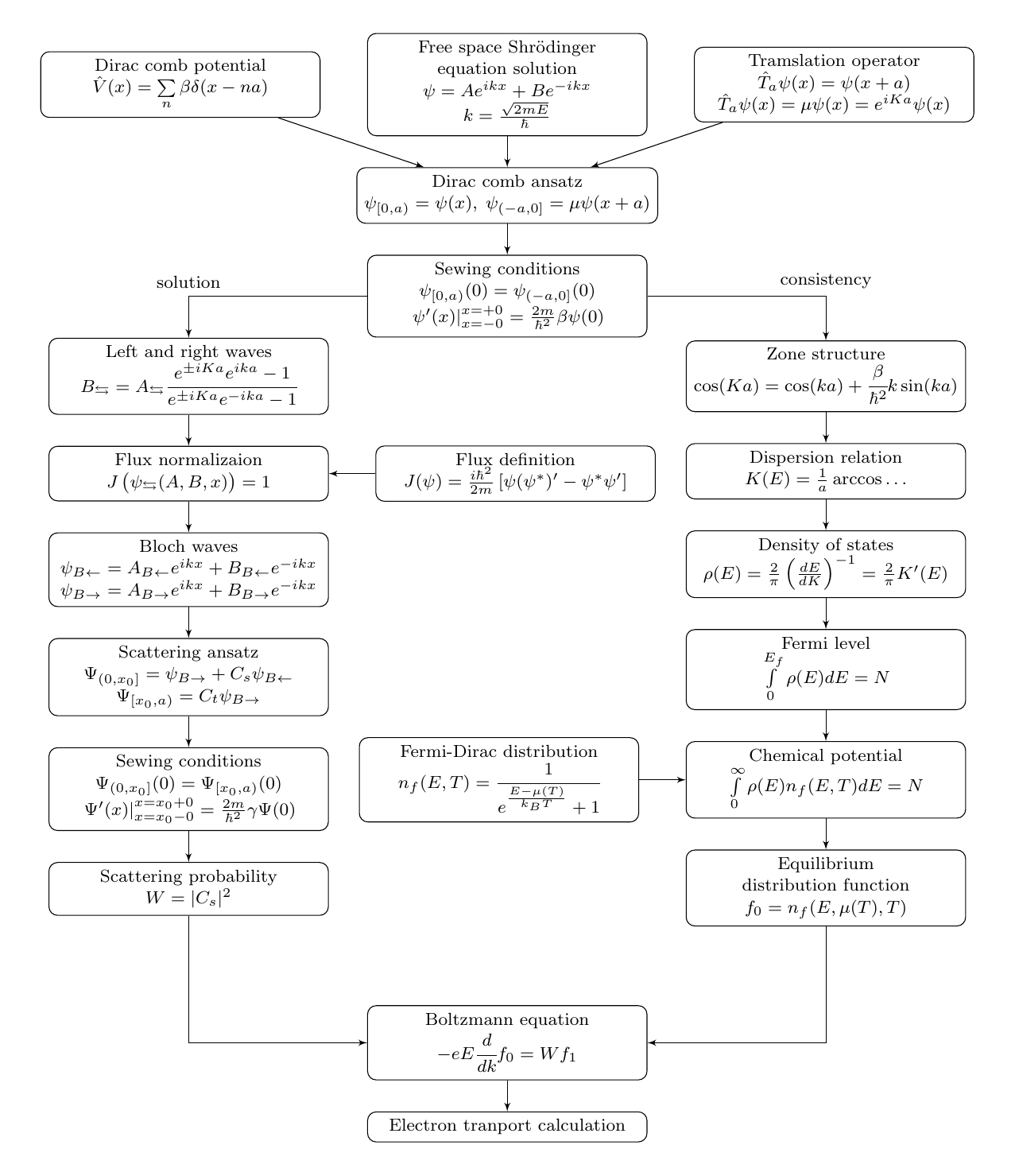}
\caption{Resistivity calculation workflow.}
\end{figure}

\end{appendices}

\end{document}